\documentclass[twoside]{article}
\usepackage{fleqn,espcrc2}

\usepackage{graphicx}
\usepackage[figuresright]{rotating}
\usepackage{cite}
\usepackage{array,dcolumn}   
\usepackage{axodraw}

\newcommand{\be}{\begin{equation}}
\newcommand{\ee}{\end{equation}}
\newcommand{\bea}{\begin{eqnarray}}
\newcommand{\eea}{\end{eqnarray}}

\newcommand{\AmS}{{\protect\the\textfont2
  A\kern-.1667em\lower.5ex\hbox{M}\kern-.125emS}}

\title{ Atomic parity violation in cesium 
and implications for the  3 - 3 - 1 models}
\author{Hoang Ngoc Long\address{{\it The Abdus Salam International 
Centre for Theoretical Physics, Trieste, Italy}}\thanks{On leave from
Institute of Physics, NCST,
P.O.Box 429, Bo Ho, Hanoi 10000, Vietnam. E-mail
address: hnlong@iop.ncst.ac.vn} and
Le Phuoc Trung\address{{\it HCMC Institute  of Physics, Mac Dinh Chi 1, 
Ho Chi Minh city, Vietnam}}}
       
\begin{document}

\begin{abstract}
 The  parity violation in cesium atom is analysed
in the framework of the models based on the 
$\mbox{SU}(3)_C\times \mbox{SU}(3)_L \times 
\mbox{U}(1)_N$ gauge group.
It is shown that in the minimal version, the main 
contribution to a deviation of weak charge 
$\Delta Q_W$ due to direct $Z'$ exchange is negative.
New data on  parity violation in the
cesium atom seems not favour to the minimal version, while
it gets a positive value in the version with 
right-handed neutrinos. We obtain 
a  bound on the $Z'$ mass at a range from
1.4 TeV to 2.6 TeV. The allowed regions for
the $Z-Z'$ mixing angle are also derived.\\
PACS number(s): 11.30.Er,12.15.Ji,12.15.Mm,12.60.Cn
\vspace{1pc}
\end{abstract}

\maketitle

\section{Introduction}
 One prediction of the standard model (SM) is the existence
of the atomic parity violation (APV).
The new data on  the APV in  cesium atom~\cite{ben} has 
caused extensive interest and reviews~\cite{view}.
Parity violation in the SM results from
exchanges of weak gauge bosons. In electron-hadron
neutral-current processes parity violation is due to
the vector axial-vector interaction in the Lagrangian.
The measurement is stated in terms of the weak charge
$Q_W$, which parameterizes the parity violating Lagrangian.
The value reported
\begin{equation}
\label{first}
Q_W ( ^{133}_{55} {\rm Cs} ) = -72.06 \pm 0.28 
({\rm expt}) \pm 0.34 ({\rm theo})
\end{equation}
represents a considerable improvement over the 
previous determination~\cite{fir}. Compared to the SM
prediction $Q_W^{SM}$, the deviation $\Delta Q_W$ is
\begin{equation}
\label{data}
\Delta Q_W \equiv Q_W({\rm Cs}) - Q_W^{\rm SM}({\rm Cs}) 
= 1.03 \pm 0.44\;,
\end{equation}
which is $2.3~ \sigma$ away from the SM prediction. This value 
has been widely used for analysis of possible new physics.

\begin{table*}[htb]
\caption{Vector and axial-vector coupling constants relevant for APV
in cesium atom for the SM and for two 3 - 3 - 1 models}
\label{table1}
\newcommand{\m}{\hphantom{$-$}}
\newcommand{\cc}[1]{\multicolumn{1}{c}{#1}}
\renewcommand{\tabcolsep}{2pc} 
\renewcommand{\arraystretch}{1.2} 
\medskip
\centering
\begin{tabular}{|c|c|c|}
\hline
Standard model & 3 - 3 - 1 model with RH neutrinos  & 
Minimal 3 - 3 - 1 model \\
\hline
\hline
$ a_e =\frac{1}{2}  $ & $a'_e =-\frac{1}{2}
\frac{1}{\sqrt{3-4 s_W^2}} $ & $ a'_e=
\frac{\sqrt{1-4 s^2_W}}{2 \sqrt{3}}\ $\\
$ v_u=\frac{1}{2}-\frac{4 s_W^2}{3}  $ & 
$ v'_u=-(\frac{1}{2}-\frac{4 s_W^2}{3})
\frac{1}{\sqrt{3-4 s_W^2}}  $ & 
$v'_u=\frac{1}{2\sqrt{3}}\frac{(-
1+6s_W^2)}{\sqrt{1-4s_W^2}}  $ \\
$ v_d=-\frac{1}{2}+\frac{2s_W^2}{3} $ & $v'_d=-
(\frac{1}{2}-\frac{s_W^2}{3})\frac{1}{\sqrt{3-4
s_W^2}}  $ & 
$v'_d=- \frac{1}{2\sqrt{3}}\frac{1}{\sqrt{1-4s_W^2}}  $\\
\hline
\end{tabular}\\[2pt]
\end{table*}

  Among the beyond SM extensions, the
recent proposed models based on the 
$\mbox{SU}(3)_C\times \mbox{SU}(3)_L \times \mbox{U}(1)_N$ gauge
group \cite{ppf,fhpp,rhnm,mpp} (hereafter 3 - 3 - 1
 models) have the following
intriguing features: firstly, the models are anomaly free only if
the number of families $N$ is a multiple of three. Further, from
the condition of QCD asymptotic freedom, which means $N < 5$, it
follows that $N$ is equal to 3. The second characteristic is that
the Lagrangians of these models possess the Peccei-Quinn
symmetry naturally, hence the strong $CP$ problem can be solved in
an elegant way \cite{pal}. The third interesting feature is that
one of the quark families is treated differently from the other
two. This could lead to a natural explanation of the
unbalancing heavy top quark in  the fermion mass
hierarchy.

There are two main versions of 
the 3 - 3 - 1 models: the {\it
minimal} model in which all lepton components $(\nu, l, l^c)_L$ of
each family belong to one and same lepton triplet and a variant,
in which right-handed (RH) neutrinos  are
included, i.e. $(\nu, l, \nu^c)_L$ (hereafter we call it the model
with right-handed neutrinos \cite{rhnm,mpp}). New gauge bosons in the
minimal model are  bileptons
 ($Y^\pm, X^{\pm\pm}$) carrying lepton number $L =\pm 2$ and a new $Z'$.
In the second model, the bileptons with lepton number $L =\pm 2$
are singly--charged $Y^\pm$ and {\it neutral} gauge bosons $X^0,
X^{*0}$ , and both are responsible for lepton--number violating
interactions.  Thus, with the present group extension there are five 
new gauge bosons and all these particles are heavy. Getting mass 
limits for these particles is one of the central tasks 
of further studies.

  We notice that it is rather easy to get bounds on masses
of the bileptons following from, for example, the ``wrong'' 
muon decay data, from the anomalous magnetic moments of the 
muon~\cite{kls} or from radiative corrections~\cite{fh,li}. 
New neutral gauge boson beyond the photon and the  $Z$ 
of the SM  has long been considered as one
of the most interesting topics. However, getting bounds on
mass of the $Z'$ in these models is rather difficult. Perhaps the 
APV is one of the most suitable subjects for this
purpose. The aim of this paper is to consider the 
atomic parity violation in the 3 - 3 - 1 models and imply it 
to get bounds on mass of the $Z'$ boson.

\section{Atomic parity violation in the minimal
3 - 3 - 1 model}

  Basic to the analysis of the experiments on parity 
violation in atoms is the electron-quark effective Lagrangian
\begin{equation}
{\cal L}_{eff}=\frac{G_F}{\sqrt{2}}(\bar{e}\gamma_\mu
\gamma_5e)(C_{1u}\bar{u}\gamma_\mu u + C_{1d} 
\bar{d}\gamma_\mu d).
\label{sta}
\end{equation}

\setcounter{figure}{0}
\begin{figure*}[t]
\centerline{\epsfxsize=12cm\epsffile{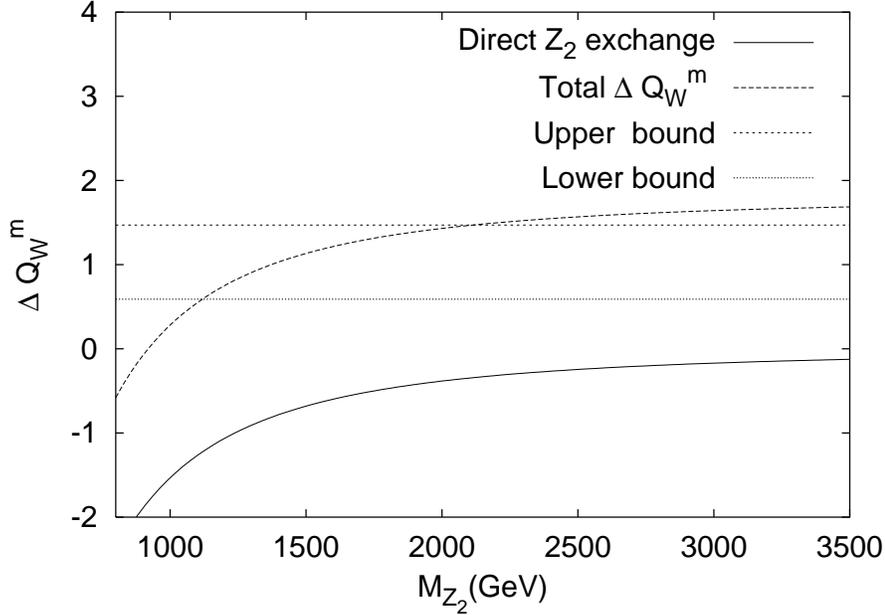}}
\caption{\label{fig0}{\em   $\Delta Q_W^m$  as 
a  function of $M_{Z'}$. The continuous line is 
obtained by neglecting the contribution from the 
$Z-Z'$ mixing. }}
\end{figure*}
\vspace*{0.3cm}

We see that it involves only electron coupling being axial 
(or spin dependent) and quark coupling vector. This interaction 
between the electron and the nucleus is thus coherent, 
proportional to the total weak charge. There is a similar 
interaction which is also generated by direct $Z$ exchange, 
but with the electron coupling being vector and the nuclear 
coupling axial vector. This kind of interaction is inhibited: 
the axial coupling to the nucleus is no longer coherent 
(only the last unpaired nucleon contributes), and the electron 
vector coupling is quite suppressed ($\sim$ 0.026). In addition, 
one knows that the electromagnetic interaction of atomic 
electrons with the nuclear anapole moment may generate a 
measurable spin dependence in atomic parity violation 
experiments. This interaction, which is considerably smaller 
than the coherent $Z$ interaction, may exceed the direct 
nuclear spin dependent parity violation for heavy 
atoms \cite{ha}. Hence one may neglect the nuclear spin 
dependent parity violations.

\setcounter{figure}{1}
\begin{figure*}[thb]
\centerline{\epsfxsize=12cm\epsffile{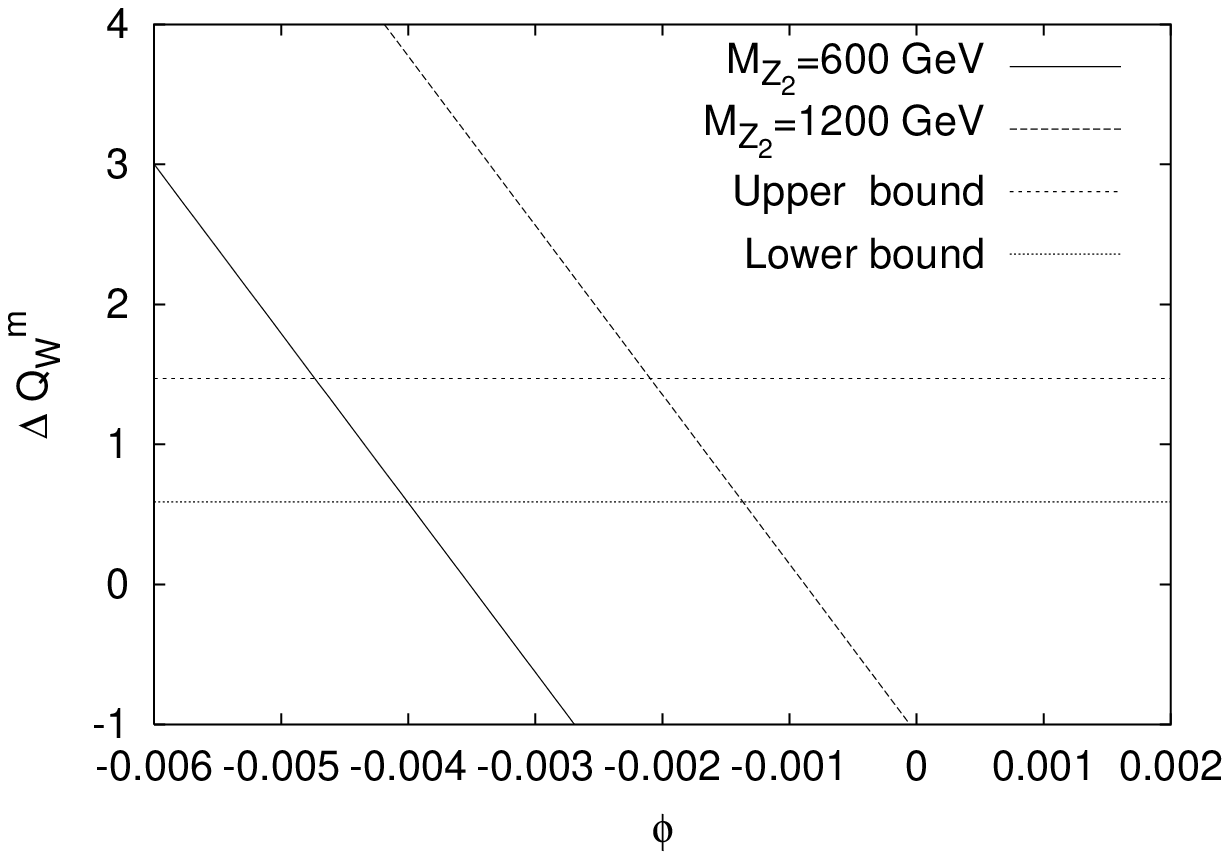}}
\caption{\label{fig1}{\em   $\Delta Q_W^m$  as 
a  function of $\phi$. The continuous line refers
to $M_{Z'}=600$ GeV, the dashed line -- 1200 GeV.}}
\end{figure*}
\vspace*{0.3cm}

  The experimental results are usually quoted in terms of the
related so-called weak charge
\begin{equation}
Q_W=-2[C_{1u}(2Z+N)+C_{1d}(Z+2N)],
\end{equation}
where $Z$ and $N$ are the number of protons and neutrons,
respectively, of the atom considered.
We are interested in the deviation $\Delta Q_W=Q_W-Q_W^{SM}$ with
respect to the SM predictions. This deviation is given as~\cite{al}
\begin{eqnarray}
\label{delta}
\Delta Q_W&=&\left[ \left(1+4\frac{s^4_W}{\cos
2\theta_W}\right)Z-N \right]\Delta \rho_M \nonumber\\
 &&+16 \left[(2Z+N)(a_e v'_u+a'_e v_u )\right.\nonumber\\
&&+\left. (Z+2 N)(a_e v'_d+a'_e v_d)
\right]\phi 
\label{dev}\\
 &&+16 \left[(2Z+N)a'_e v'_u\right.\nonumber\\
&&\left.+(Z+2N)a'_e
v'_d\right]\frac{m^2_{Z}}{M^2_{Z'}},\nonumber
\end{eqnarray}
where $\phi$ is the mixing angle between the new gauge boson $Z'$ 
and the canonical $Z$, $v_f$, $a_f$,
$v'_f$, $a'_f$ are couplings of $Z$, $Z'$ to fermions. 
The couplings are defined in the neutral interactions
\begin{eqnarray}
{\cal L}(Z)&=&\frac{g}{2\cos \theta_W}\sum_f(v_f
\bar{f}\gamma_{\mu}f+a_f \bar{f}\gamma_{\mu}\gamma_5f)Z^{\mu},
\nonumber\\ 
{\cal L}(Z')&=&\frac{g}{2\cos\theta_W}\sum_f(v'_f
\bar{f}\gamma_{\mu}f+a'_f \bar{f}\gamma_{\mu}\gamma_5f)Z'^{\mu},
\nonumber
\end{eqnarray}
where the summation is over all relevant fermions. 
The coupling constants relevant for calculations of
APV in the cesium atom for the SM, 3 - 3 - 1 model 
with RH neutrinos and the minimal version are presented 
in Table 1, in which the
usual notation is used: $s_W\equiv \sin \theta_W$. More details on 
the 3 - 3 - 1 models can be found in Refs.~\cite{ppf,rhnm,dng}.

  We notice that the first term on the right-handed side of 
Eq. (\ref{dev}) is
proportional to $\Delta \rho_M$, which is an additional 
contribution to the $\rho$ parameter arising from the mixing. 
This term receives two independent contributions. The first one 
comes from the overall
$\rho$ factor on the right-handed side of Eq. (2.15) in
\cite{al}. The second one arises from a shift in $\sin^2 \theta_W$, 
which is proportional to $\Delta \rho_M$. The second term is due 
to the $Z-Z'$ mixing.  These two terms are ultimately
dependent upon the mixing angle $\phi$ and particle content. 
The third term, which is 
proportional to the squared mass ratio $m_Z^2/M_{Z'}^2$, 
is due to the direct $Z'$ exchange.

  In the 3 - 3 - 1 models, deviation of
the parameter $\Delta \rho_M$ gets contributions
from both oblique corrections and the
$Z-Z'$ mixing, and it relates to the
oblique radiative parameter  $T$ ~\cite{ta} as 
the following
\begin{equation}
\Delta \rho_M= \alpha T.
\end{equation}

  In the minimal model, it was shown that~\cite{fh,li}
\begin{eqnarray}
 T^{ m}&=&\frac{3\sqrt{2}G_F}{16\pi^2\alpha}
\left[M_{++}^2 + M_+^2\right.\nonumber\\
&&\left. - \frac{2M_{++}^2M_+^2}{M_{++}^2-M_+^2}
  \ln \frac{M_{++}^2}{M_+^2}
\right] \nonumber \\
       & &+\frac{1}{4\pi\ s^2_W}
\left[2\bar{F}_0(0,M_{++},M_+)\right. \nonumber\\
&&+3\left.
t^2_W \ln\frac{M_{++}^2}{M_+^2}
\right]\nonumber\\
&&+ \frac{\tan^2 \phi}{\alpha} \left( \frac{
M^2_{Z'}}{m^2_{Z}} - 1 \right),
\label{ttotm}
\end{eqnarray}
where ${\bar F}_0(s,M,m)\equiv \int_0^1 dx \ln
[(1-x)M^2 +xm^2-x(1-x)s] -\ln Mm $, and
$M_+, M_{++}$ stand for masses of the singly-
and doubly-charged gauge bosons
$Y^+, X^{++}$, respectively.  
The first two terms in 
(\ref{ttotm}) are polynomial in the ratio between 
mass splitting and the bilepton  mass~\cite{li}:
 $\varepsilon \equiv \varepsilon(M_+,M_{++}) = \frac{M_+^2 -
M_{++}^2}{M_{++}^2}$. The superscript $ m$ in $T^{m}$
refers to the parameter in the minimal version.

  By the spontaneous symmetry breaking, the bilepton mass
splitting is given (see N. A. Ky {\it et al} in~\cite{kls})
\be
|M_{++}^2 - M_+^2| \leq 3~ m^2_W.
\label{msip}
\ee
Therefore the parameter $\varepsilon$ is bounded too. 
This leads to  limitation in the $T$ parameter.

  With the $Z'$  mass in a range of 1 TeV scale, the 
last term in (\ref{ttotm}) gives just contribution of
about 4\% to the $T^{m}$ parameter~\cite{li}. 

  The $Z-Z'$ mixing angle $\phi$ is constrained~\cite{dng}:
 $ -5 \times 10^{-3} \leq \phi \leq 7\times
10^{-4}$, hence the two last terms in (\ref{dev}) give
a dominant contribution. 

  Using the following data~\cite{pdg}: 
$s_W^2 = 0.23117, m_Z = 91.1882 {\rm GeV}, M_+ = 230 \
 {\rm GeV}, M_{++}=300\  {\rm GeV}$, we obtain
\be
\Delta Q_W^m \simeq - 0.01 - 1210.07 \phi - 184.32 
\left(\frac{m_Z^2}{M_{Z'}^2}\right).
\label{detm}
\ee 

  In Fig. 1 we plot the 
direct $Z'$ exchange contribution 
 (the continuous line) and  
$\Delta Q_W^m$ (the dashed line) as a function of 
$M_{Z'}$. 
We see that the direct $Z'$ exchange contribution
is negative. The coefficient of $\phi$ in the
second term - the $Z-Z'$ mixing contribution is 
negative too. Therefore the total 
$\Delta Q_W^m$ get a negative value if $\phi$ is positive.
This circumstance  is excluded at 99\% CL (see R. 
Casalbuoni {\it et al.} in~\cite{view}).
The dashed line corresponds to the total
$\Delta Q_W^m$ at the $\phi = - 1.5 \times 10^{-3}$.
In this case the constraint for  the $Z'$ mass is
$1.15\ {\rm TeV} \le M_{Z'} \le  2.2 $  TeV. 

\setcounter{figure}{2}
\begin{figure*}[t]
\centerline{\epsfxsize=12cm\epsffile{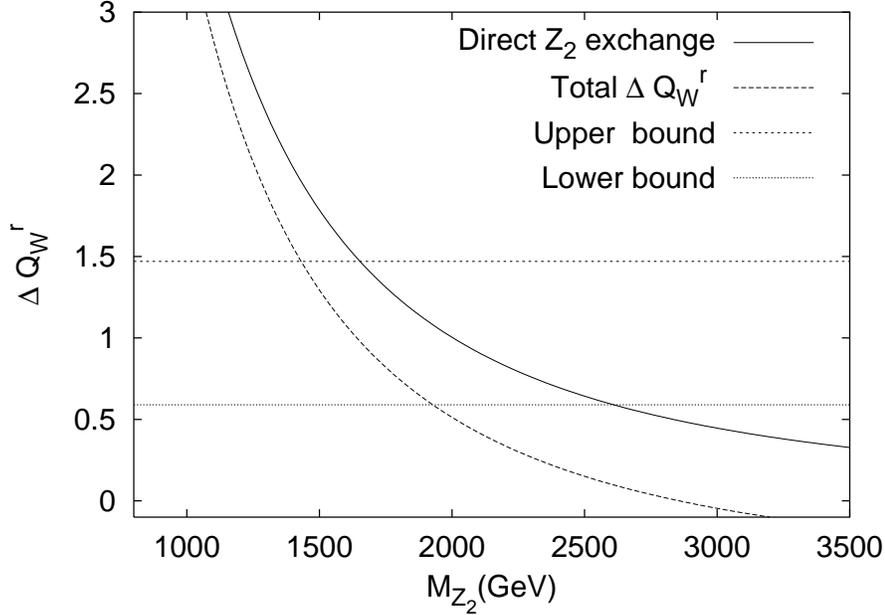}}
\caption{\label{fig2}{\em   $\Delta Q_W^r$  as 
a  function of $M_{Z'}$.
The continuous line is 
obtained by neglecting the contribution from the 
$Z-Z'$ mixing. }}
\end{figure*}
\vspace*{0.3cm}

  In Fig. 2 we plot $\Delta Q_W^m$ as a function
of the $Z-Z'$ mixing angle $\phi$ at two values for $M_{Z'}=600$
GeV (the dashed line) and  $M_{Z'} = 1200$ GeV (the 
continuous line). However we can make 
the $\Delta Q_W^m$ positive  by taking
$\phi$ is negative and large enough, say in order
of few $\times 10^{-3}$. 
In conclusion, to be consistent with the new data on 
parity violation in the cesium atom, the $Z-Z'$ mixing angle 
$\phi$ in the minimal 3 - 3 - 1 model has to be negative and is
in order of a few  $\times 10^{-3}$
\bea
-0.0045 &\leq & \phi \leq -0.004\    
{\rm for}\  M_{Z'} = 600\ {\rm GeV},\nonumber\\
 -0.002 &\leq & \phi \leq -0.0018\    
{\rm for}\ M_{Z'} = 1200\ {\rm GeV}.\nonumber
\eea

\section{Atomic parity violation in the 
3 - 3 - 1 model with RH neutrinos}

  In the model with RH neutrinos the bileptons 
$ Y^+, X^0$ make an $SU(2)_L$ doublet with
hypercharge $Y=\frac 1 2$. The $T$ parameter
in this model is given as~\cite{li}
\begin{eqnarray}
 T^{ r}& = &   \frac{3\sqrt{2}G_F}{16\pi^2\alpha}
\left[ 
  M_+^2 + M_0^2 - \frac{2M_+^2M_0^2}{M_+^2-M_0^2}
  \ln \frac{M_+^2}{M_0^2}
\right] \nonumber \\
       &  & + \frac{1}{4\pi\ s^2_W}
\left[ 2    \bar{F}_0(0,M_+,M_0) +  \ 
t^2_W \ln\frac{M_+^2}{M_0^2}
\right]\nonumber\\
&&+ \frac{\tan^2 \phi}{\alpha} \left( \frac{
M^2_{Z' }}{m^2_{Z  }} - 1 \right),
\label{ttot} 
\end{eqnarray}
where $M_+, M_{0}$ stand for masses of the singly-charged
and neutral gauge bosons $Y^+, X^{0}$, respectively.
Again, the first two terms in (\ref{ttot}) are polynomial
in $\epsilon \equiv \epsilon(M_+,M_0) = \frac{M_+^2 -
M_0^2}{M_0^2}$.

\setcounter{figure}{3}
\begin{figure*}[thb]
\centerline{\epsfxsize=12cm\epsffile{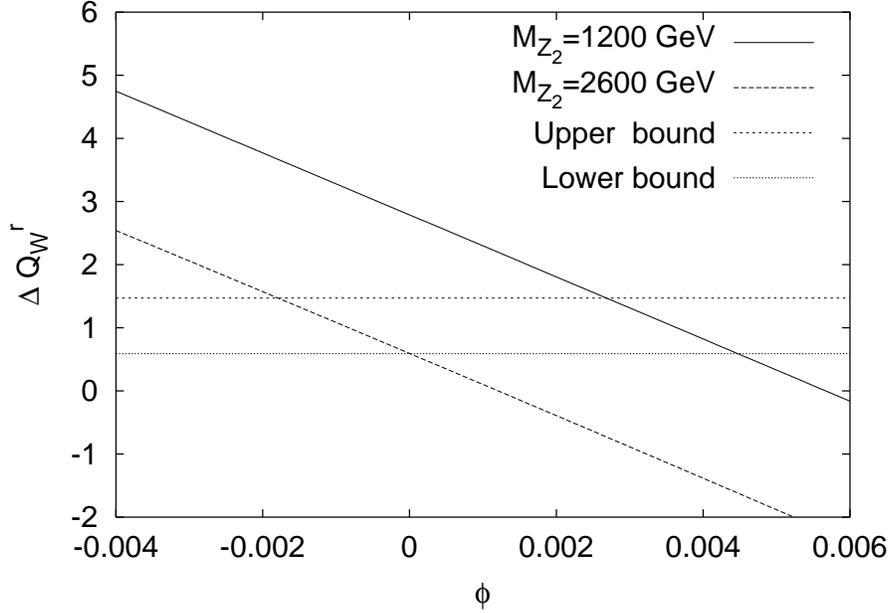}}
\caption{\label{fig3}{\em   $\Delta Q_W^r$  as 
a  function of $\phi$.  The continuous line refers
to $M_{Z'}=1200$ GeV, the dashed line -- 2600 GeV.}}
\end{figure*}
\vspace*{0.5cm}

  As before, the spontaneous symmetry breaking gives 
the bilepton mass splitting~\cite{li}
\be
|M_{0}^2 - M_+^2| \leq  m^2_W.
\label{ms2}
\ee
Hence the parameter $\epsilon$ is bounded shorter
than one in the minimal version. The contribution  
from the first term in (\ref{dev}) is negligible.

  The $Z-Z'$ mixing angle $\phi$ is also constrained to be 
very small $ -2 \times 10^{-4} \leq \phi \leq 3\times
10^{-3}$, therefore the last term in (\ref{dev}) gives
a dominant contribution. 

  Taking into account that~\cite{li}: $M_+ = 230 
\ {\rm GeV}, M_{0}=250 \ {\rm GeV}$, we get
\be 
\Delta Q_W^r \simeq - 0.00092 - 490.71 \phi + 482.99 
\left(\frac{m_Z^2}{M_{Z'}^2}\right).
\label{detr}
\ee 
Here the sign of the last term is plus.  
The coefficient of $\phi$ 
in the second term of $\Delta Q_W^r$, where $ r$
refers to the model with RH neutrinos, is negative
and is much smaller than those in the minimal version.
\vspace*{0.5cm}

  In Fig. 3 we plot the direct $Z'$ exchange contribution 
(the continuous line) and  
$\Delta Q_W^r$ (the dashed line) as a function of $M_{Z'}$. 
Here we take the $Z-Z'$ mixing angle $\phi = 0.001$.
The contribution of the second 
term is negative if the mixing angle $\phi$ is
positive. However it is just 10\% of total value.
From the figure we get  the constraint for the $Z'$ mass:
$1.4\ {\rm TeV} \le M_{Z'} \le 1.9\ {\rm TeV}$
 in the case of the total $\Delta Q_W^r$. 
This number is slightly higher:
$1.7\ {\rm TeV} \le M_{Z'} \le 2.6\ {\rm TeV}$
 in the case of neglecting the $Z-Z'$ mixing.

  In Fig. 4 we plot $\Delta Q_W^r$ as a function
of the $Z-Z'$ mixing angle $\phi$ at two values for $M_{Z'}=1200$
GeV (the dashed line) and  $M_{Z'}=2600$ GeV (the 
continuous line). From the figure it follows that an allowed 
region for the mixing angle is
\bea
0.0028 &\leq&  \phi \leq 0.0046\ \   
{\rm for}\ M_{Z'} = 1200\ \  {\rm GeV},\nonumber\\
- 0.0018 &\leq & \phi \leq 0\ \   
{\rm for}\ M_{Z'} = 2600\ \  {\rm GeV}.\nonumber
\eea
 
Summarizing this section, we conclude that the APV data
provides a large room for the 3 - 3 - 1 model with 
RH neutrinos. 

\section{Conclusions}

  We have considered  parity violation in the cesium atom.
It is shown that in the minimal version the main 
contribution to a deviation of weak charge 
$\Delta Q_W$ due to direct $Z'$ exchange,  is {\it negative}.
To make the deviation positive, the $Z-Z'$ mixing angle
$\phi$ has to be {\it negative} and large enough, say in order
of few  $\times 10^{-3}$. In the case of
$\phi = - 0.0015$ the constraint for 
the $Z'$ mass is 
$1.15\ {\rm TeV} \le M_{Z'} \le 2.2\ {\rm TeV}$.
New data on  parity violation in the 
cesium atom seems not favour to the minimal version.

The deviation of weak charge  in the 3 - 3 -1 model
with RH neutrinos  is {\it positive}
 and is  mainly  from the $Z'$ exchange. 
For the mixing angle $\phi = 0.001$
 we get  the constraint for the $Z'$ mass:
$1.4\ {\rm TeV} \le M_{Z'} \le 1.9\ {\rm TeV}$
 in the case of the total $\Delta Q_W^r$. 
Neglecting the $Z-Z'$ mixing we get
the constraint:
$1.7\ {\rm TeV} \le M_{Z'} \le 2.6\ {\rm TeV}$.
If mass of the $Z'$ lying around present experimental
limit of about 600 GeV~\cite{exp}, then  the allowed region 
for the $Z-Z'$ mixing angle in the minimal version is:
$-0.0045 \leq  \phi \leq - 0.004$, which is comparable to
limit extracted from the $Z$-decay experiments.
For the model with RH neutrinos, in order to have a mixing 
angle consistent to that from the $Z$ decay data, i.e.
$\phi \approx -2 \times 10^ {-4} \div 3 \times 10^{-3}$,
mass of the $Z'$ has to be heavier than 1 TeV.

  Finally, we mention that  parity violation
in the cesium atom provides a useful tool for 
getting bounds on mass of the new $Z'$ and its mixing
with the Z boson in the SM. 
 
{\bf Acknowledgments}

  One of the authors (H. N. L.) would like to thank the Abdus Salam
International Centre for Theoretical Physics, Trieste, Italy for financial 
support and hospitality. This work was supported in part
by the Natural Science Council of Vietnam.


\begin{thebibliography}{99}
\bibitem{ben} S. C. Bennett, C. E. Wieman, Phys. Rev. Lett.
 82 (1999) 2484.
\bibitem{view} R. Casalbuoni, S. De Curtis, D. Dominici and
R. Gatto, Phys. Lett. B  460 (1999) 135 ;
J. L Rosner, Phys. Rev. D  61 (2000) 016006;
V. Barger and K. Cheung, Phys. Lett.  B  480 (2000) 149;
J. Erler and P. Langacker, Phys. Rev. Lett.  84 (2000) 212.
\bibitem{fir} C. S. Wood, {\it et al.,} Science  275 
(1997) 1759. 
\bibitem{ppf} F. Pisano and V. Pleitez, Phys. Rev. D  46
(1992) 410; P. H. Frampton, Phys. Rev. Lett.   69 (1992) 2889.
\bibitem{fhpp} R. Foot, O.F. Hernandez, F. Pisano and
V. Pleitez, Phys. Rev. D  47 (1993) 4158.
\bibitem{rhnm}R. Foot, H. N. Long and Tuan A. Tran,
Phys. Rev. D   50 (1994) R34; H. N. Long, Phys. Rev. D  
 53 (1996) 437; Phys. Rev. D 54 (1996) 4691.
\bibitem{mpp}J. C. Montero, F. Pisano and V. Pleitez, Phys. Rev.
D  47 (1993) 2918.
\bibitem{pal}P. B. Pal, Phys. Rev. D  52 (1995) 1659.
\bibitem{kls} H. Fujii, S. Nakamura and K. Sasaki,
Phys. Lett. B  299 (1993) 342;
N. A. Ky, H. N. Long and D. V. Soa, 
Phys. Lett.  B  486 (2000) 140.
\bibitem{fh} K. Sasaki, Phys. Lett. B  308 (1993) 297;
P. H. Frampton and M. Harada,
Phys. Rev. D  58 (1998) 09513.
\bibitem{li} H. N. Long and T. Inami, Phys. Rev. D  61 
 (2000) 075002.
\bibitem{ha} W. C. Haxton, Science  275
 (1997) 1753. 
\bibitem{al} G. Altarelli,  R. Casalbuoni, S. De Curtis, 
N. Di Bartolomeo, F. Feruglio and
R. Gatto, Phys. Lett. B  261 (1991) 146.
\bibitem{dng}D. Ng, Phys. Rev.  D  49 (1994) 4805.
\bibitem{ta} M. E. Peskin and T. Takeuchi, Phys. Rev.
Lett.  65 (1990) 964; Phys. Rev, D  46 (1992) 381.
\bibitem{pdg} D. E. Groom {\it et al}., Particle Data Group,
Eur. Phys. J. C  15 (2000) 1.
\bibitem{exp} CDF Collaboration, F. Abe {\it et al}.,
Phys. Rev. Lett.  79 (1997) 2192.
\end{thebibliography}
\end{document}